\documentclass[10pt,conference]{IEEEtran}
\IEEEoverridecommandlockouts
\usepackage{cite}
\usepackage{amsmath,amssymb,amsfonts}
\usepackage{algorithmic}
\usepackage{graphicx}
\usepackage{textcomp}
\usepackage[caption=false,font=footnotesize]{subfig}
\usepackage{xcolor}
\usepackage{hyperref}
\newcommand{\Autoref}[1]{%
  \begingroup
  \def\sectionautorefname{Section}%
  \def\subsectionautorefname{Section}%
  \def\subsubsectionautorefname{Section}%
  \autoref{#1}%
  \endgroup
}
\usepackage[most]{tcolorbox}
\usepackage{booktabs}
\usepackage{soul}
\def\BibTeX{{\rm B\kern-.05em{\sc i\kern-.025em b}\kern-.08em
    T\kern-.1667em\lower.7ex\hbox{E}\kern-.125emX}}

\definecolor{findingbg}{gray}{0.93}
\definecolor{findingbar}{RGB}{45,52,63}

\newtcolorbox{findingbox}{
enhanced,
width=\linewidth,
colback=findingbg,
colframe=findingbg,
boxrule=0pt,
arc=0pt,
outer arc=0pt,
left=9pt,
right=8pt,
top=5pt,
bottom=5pt,
borderline west={3pt}{0pt}{findingbar},
fontupper=\normalfont,
before skip=6pt,
after skip=6pt
}

\begin{document}

\title{
Agents with Feelings? Personality and Emotion in Multi-Agent Software Teams
}

\author{\IEEEauthorblockN{1\textsuperscript{st} Yunyan Ding}
\IEEEauthorblockA{\textit{Department of Computer Science} \\
\textit{University of California, Irvine}\\
Irvine, United States \\
yunyad1@uci.edu}
\and
\IEEEauthorblockN{2\textsuperscript{nd} Thomas Zimmermann}
\IEEEauthorblockA{\textit{Department of Informatics} \\
\textit{University of California, Irvine}\\
Irvine, United States \\
tzimmer@uci.edu}
\and
\IEEEauthorblockN{3\textsuperscript{rd} Iftekhar Ahmed}
\IEEEauthorblockA{\textit{Department of Informatics} \\
\textit{University of California, Irvine}\\
Irvine, United States \\
iftekha@uci.edu}
}

\maketitle

\begin{abstract}

Multi-agent LLM systems for Software Engineering (SE) typically differentiate agents through roles and workflows, but little is known about how agents' behavioral profiles affect team performance. We investigate the impact of personality and emotion profiles on LLM agent teams using a psychology-informed framework that combines Big Five personality traits, basic emotions, SE-relevant work styles, and task roles. We evaluate 78 team-profile configurations across code generation and code review using four LLMs and 659 task instances. Results show that profile choice substantially affects both performance and team behavior. For code generation, the gap between the best and worst shared-profile configurations reaches 7.1--11.3 percentage points in pass@1 across models, while the best mixed-profile configuration outperforms the best shared-profile configuration in six of eight model--task settings. Profiles also influence collaboration dynamics and cost: fear and high-conscientiousness profiles increase revision activity, over-revision, and token usage without consistent performance gains. These findings identify agent profiles as an important design dimension in multi-agent SE systems, affecting not only task outcomes but also the efficiency of collaboration. 

\end{abstract}


\section{Introduction}
\label{sec:introduction}

Large Language Models (LLMs) are increasingly used to support a wide range of Software Engineering (SE) tasks~\cite{fan2023large,hou2024large}. To address the collaborative nature of software development, recent SE-oriented multi-agent systems, such as ChatDev and MetaGPT~\cite{qian2024chatdev,hong2024metagpt}, organize LLM agents around software development workflows and assign them roles that mirror those in software teams. These systems primarily differentiate agents by their responsibilities, defining what each agent does. %

However, in human software teams, individuals differ not only in the work they perform but also in how they approach it, due to characteristics such as personality, emotional state, and work style. Psychology research has shown that personality traits are associated with individual job performance~\cite{barrick1991big,zell2022big}. At the team level, personality composition influences team effectiveness and performance~\cite{peeters2006personality,bell2007deep}. SE research similarly suggests that personality and personality diversity affect team outcomes~\cite{pieterse2006software,karn2006follow,acuna2015team}. Beyond relatively stable personality traits, transient emotional states also influence behavior, decision-making, and performance~\cite{weiss1996affective}. These findings raise the question of whether personality and emotion profiles similarly shape the behavior and performance of LLM-agent teams.

Prior research has shown that persona-based prompts can influence LLM behavior~\cite{jiang2023evaluating,serapio2023personality} and can steer LLM outputs toward specific roles or human subpopulations~\cite{argyle2023out}. Similarly, emotion-oriented prompting studies indicate that emotional cues can influence LLM behavior and performance~\cite{li2023large}. In SE, persona-guided prompting has been shown to improve code generation in single-agent settings~\cite{guo2025personality}. Beyond SE, studies of multi-agent LLM systems have found that assigning agents different personality traits, such as Big Five-based traits, can influence team performance~\cite{duan2025power}. However, while prior work has examined personality- and emotion-based prompting in individual LLMs and non-SE multi-agent settings, their impact on collaborative SE teams remains largely unexplored. Understanding this relationship can provide insights into how agent profiles should be designed in multi-agent SE systems.

To address this gap, we conduct a systematic study of how personality and emotion profiles influence the performance and behavior of LLM teams on collaborative SE tasks. 
We refer to a personality and emotion configuration assigned to an agent as a \textit{profile}, and to the corresponding natural-language description that reflects a profile, a specific SE task role, and work-style tendencies as a \textit{persona description}.

We investigate two profile settings: \textbf{\emph{shared-profile teams}}, in which all agents use the same profile, and \textbf{\emph{mixed-profile teams}}, in which agents are assigned different profiles. We evaluate these settings across four LLMs representing multiple model families and capability levels. 

Because a systematic study of profile effects requires a principled approach to profile construction, we develop a \textbf{\emph{psychology-informed framework}} that integrates 
\begin{itemize}
\item Big Five personality traits~\cite{soto2017next,mccrae1992introduction} (i.e., conscientiousness, openness, and extraversion), 
\item SE-relevant work styles from the O*NET taxonomy~\cite{ONETContentModel, handel2016net} (i.e., innovation, adaptability, achievement orientation, intellectual curiosity, attention to detail, cautiousness, and tolerance for ambiguity as well as their opposing behavioral poles), 
\item emotion categories grounded in basic emotion theory and SE emotion research~\cite{ekman1992argument,sanchez2019taking} (i.e., anger, fear, disgust, sadness, neutral, and happiness), and 
\item task roles (i.e., planner, implementer, and reviewer for code generation; writer and supervisor for code review). 
\end{itemize}
We treat personality, emotion, and task role as experimental factors, and derive work styles from personality profiles to construct coherent persona descriptions.

Using this framework, we study two common SE tasks: \emph{code generation} and \emph{code review}. We select code generation as a widely studied application of LLMs in SE and as a direct measure of a team's ability to produce executable solutions~\cite{jiang2026survey,fan2023large,hou2024large}. We select code review as a complementary activity in which LLMs evaluate software artifacts and provide natural language feedback~\cite{ramesh2025automated,ren2025hydra}. Together, these tasks cover both code-producing and code-evaluating forms of LLM-based SE work. We address the following three research questions:

\textbf{RQ1:} How do different shared-profile configurations affect the performance of LLM teams?

\textbf{RQ2:} How do mixed-profile and shared-profile team configurations differ in their impact on LLM team performance?

\textbf{RQ3:} How do personality and emotion profiles influence coordination and collaboration behaviors within LLM teams?

In summary, our study makes the following contributions:
\begin{itemize}
    \item We propose a psychology-informed framework for constructing persona descriptions for LLM agents in software engineering tasks by integrating personality traits, SE-relevant work styles, emotions, and task roles.
\item We investigate how personality and emotion profiles influence LLM team performance and collaborative behavior, and how profile assignment strategies (shared-profile versus mixed-profile teams) affect task outcomes.

\item We conduct a large-scale empirical study of 78 agent-profile configurations across two software engineering tasks, including 54 shared-profile configurations and 24 mixed-profile configurations, using four LLMs spanning three model families and four scales.

\end{itemize}

The remainder of this paper is structured as follows. \Autoref{sec:related-work} reviews related work. \Autoref{sec:methodology} describes our profile-construction framework and methodology for the two selected SE tasks. \Autoref{sec:experimental-setup} outlines the experimental setup, including models, datasets, metrics, and implementation details. \Autoref{sec:results} presents the results for our research questions. \Autoref{sec:discussion} discusses the implications of our findings. \Autoref{sec:threats} discusses threats to validity. \Autoref{sec:conclusion} concludes the paper.

\section{Related Work}
\label{sec:related-work}

\subsection{Multi-Agent LLM Systems in SE}

Multi-agent LLM systems have become a common approach for organizing complex SE tasks. These systems decompose development activities across multiple agents and coordinate their interactions through structured workflows. For example, ChatDev models software development as a virtual software company in which agents assume different development roles~\cite{qian2024chatdev}, while MetaGPT organizes collaboration through role-specific procedures and intermediate artifacts~\cite{hong2024metagpt}. More general frameworks, such as AutoGen, CAMEL, and AgentVerse, support collaboration through role instructions, shared context, and multi-step interactions~\cite{wu2024autogen,li2023camel,chen2024agentverse}. Collectively, these systems establish role assignment and coordination as key design dimensions of LLM teams. However, they primarily differentiate agents through their assigned responsibilities. In contrast, we investigate agent profiles as an additional design dimension within role-based LLM teams rather than proposing a new coordination architecture.

\subsection{Personality, Emotion, and Personas in Human Teams and LLM Agents}

Human software developer teams differ not only in assigned responsibilities but also in personality traits, affective states, and work styles. Psychology research has shown that personality traits are associated with individual job performance~\cite{barrick1991big,zell2022big}, while team personality composition is linked to team effectiveness~\cite{peeters2006personality,bell2007deep}. SE studies similarly suggest that personality and personality diversity influence software team outcomes~\cite{pieterse2006software,karn2006follow,acuna2015team}. Emotions also play an important role in software development work. Prior research has shown that workplace experiences can shape how people feel and behave~\cite{weiss1996affective}, and SE studies have examined developers' emotions and sentiment during software development and communication~\cite{sanchez2019taking,lin2018sentiment}. Together, these findings suggest that team outcomes depend not only on assigned responsibilities but also on how team members approach their work.

Recent LLM research suggests that personality- and emotion-related characteristics can be induced through prompting. Prior studies have examined personality-oriented prompts and their effects on LLM behavior~\cite{jiang2023evaluating,serapio2023personality}, while persona prompts have been used to guide outputs toward specific roles, perspectives, or user populations~\cite{argyle2023out}. Emotion-oriented prompting has similarly been shown to influence LLM responses~\cite{li2023large}. Within software engineering, personality-guided prompting has been studied for code generation~\cite{guo2025personality}, and persona- and personality-based approaches have been used to diversify LLM-driven testing behavior~\cite{yu2026towards,chen2025mimic}. Beyond SE, personality-based agent configurations have been shown to influence multi-agent collaboration~\cite{duan2025power}, while persona-driven agents have been used in social simulation~\cite{park2023generative}.

Despite these advances, the effects of personality and emotion profiles in multi-agent SE systems remain largely unexplored. It remains unclear how personality and emotion profiles influence the performance and behavior of LLM teams on collaborative SE tasks, whether mixed-profile teams differ from shared-profile teams, and how profile assignments shape team interactions. Our work addresses this gap through a systematic study of psychology-informed agent profiles in multi-agent SE workflows.

\subsection{LLM-Based Code Generation}

Code generation is one of the most widely studied applications of LLMs in software engineering. Benchmarks such as HumanEval, MBPP, and LiveCodeBench evaluate generated solutions using execution-based measures of functional correctness~\cite{chen2021evaluating,austin2021program,jain2025livecodebench}. Prior work has explored a broad range of approaches, including model design and training data~\cite{nijkamp2022codegen,li2023starcoder,roziere2023code}, prompting strategies and LLM-based coding techniques~\cite{fan2023large,hou2024large,jiang2026survey}, feedback-driven repair~\cite{shinn2023reflexion,madaan2023self}, and agent-based workflows~\cite{huang2023agentcoder,islam2024mapcoder}. These studies have established code generation as a standard setting for evaluating LLM-based software engineering systems. In our work, we use code generation as an execution-based task that enables direct evaluation of team outputs through functional correctness.

\subsection{LLM-Based Code Review}

Code review is a central software engineering practice in which developers inspect code changes and provide feedback~\cite{bacchelli2013expectations}. Such feedback often addresses correctness, maintainability, readability, and project-specific conventions. LLM-based automated code review extends this practice by generating natural-language feedback on code changes. Prior work has studied pre-trained models for review-comment generation~\cite{li2022automating}, LLM-based automated reviewers~\cite{ramesh2025automated}, and multi-agent review systems such as Hydra-Reviewer~\cite{ren2025hydra}. We use code review as a complementary reference-based task in which generated feedback is compared against human-written review comments. Together with code generation, this task enables us to study personality and emotion profile effects under both execution-based and reference-based evaluation settings.

\section{Methodology}
\label{sec:methodology}

\begin{figure}[tbp]
    \centering
    \includegraphics[width=0.85\linewidth]{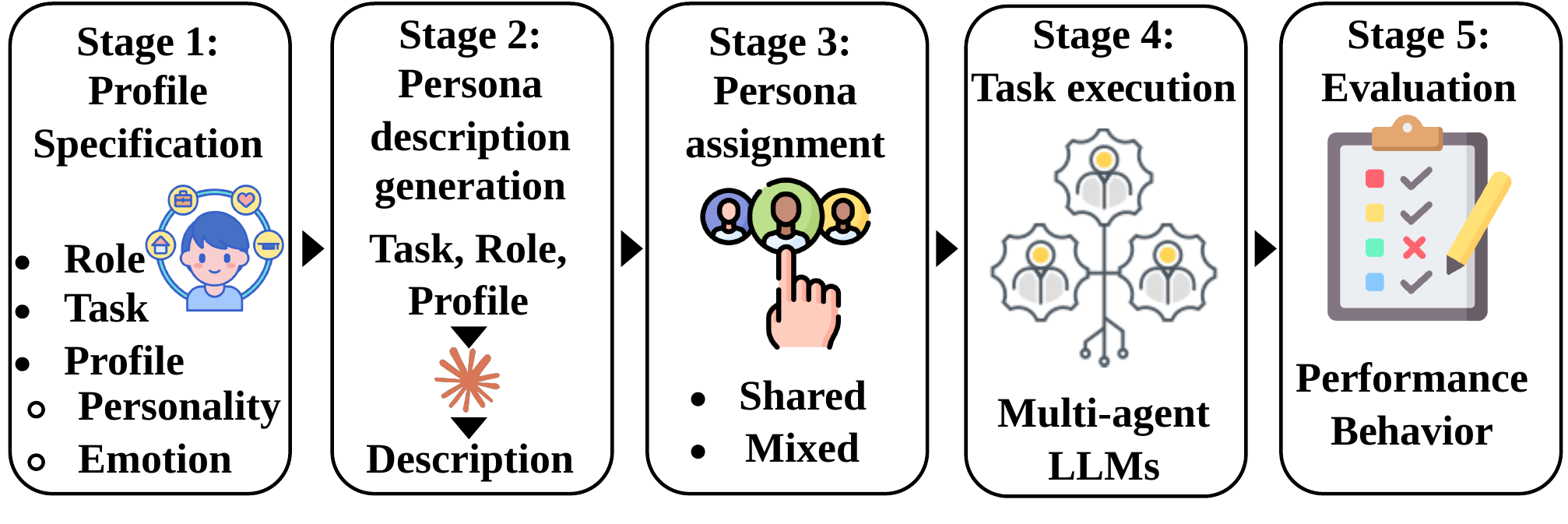}
    \caption{
    Overview of our study design, including profile specification, persona description generation, persona assignment, multi-agent task execution, and evaluation of performance and team behavior.
    }
    \label{fig:overview}
\end{figure}

In this section, we describe the persona-construction framework and the multi-agent workflows for the two tasks, corresponding to Stages 1, 2, and 4 in \autoref{fig:overview}. The profile-assignment strategy and evaluation setup, corresponding to Stages 3 and 5, are described in \Autoref{sec:experimental-setup}.

\subsection{Psychology-Informed Persona Framework} 
\label{sec:persona-framework}

We construct role-specific persona descriptions from two psychology-informed components, \textit{personality} and \textit{emotion}, and derive SE-relevant  \textit{work styles} to adapt these components to task-specific roles.
\textit{Personality} defines the agent's stable disposition\cite{mccrae1992introduction}, \textit{emotion} specifies its temporary affective state\cite{scherer2005emotions}, and \textit{work styles} reflect work-related tendencies in how the agent approaches software-development tasks\cite{ONETContentModel,NAP12814}.

\paragraph{Personality}
We adopt the Five-Factor Model as the basis for personality~\cite{mccrae1992introduction,soto2017next}, which characterizes personality along five dimensions: Conscientiousness, Agreeableness, Neuroticism, Openness, and Extraversion. Because our goal is to study personality profiles that are likely to influence SE task performance, we focus on Conscientiousness, Openness, and Extraversion, as prior work~\cite{gnambs2015makes} has reported reliable associations between these traits and programming aptitude. In contrast, no such association has been consistently observed for Agreeableness or Neuroticism~\cite{gnambs2015makes}. Each selected trait is assigned one of two levels, \textsc{High} or \textsc{Low}. A personality profile is therefore defined by a combination of trait levels (e.g., \textsc{Low Conscientiousness}, \textsc{High Openness}, and \textsc{Low Extraversion}), yielding $2^3 = 8$ possible personality profiles. To provide a baseline, we also include an all-\textsc{Neutral} profile in which all three traits are described as balanced, resulting in nine personality profiles in total.

\paragraph{Emotion}

We select emotions based on both basic emotion theory and empirical studies of developer affect in SE. A systematic literature review reports that anger, fear, disgust, sadness, joy, love, and happiness are among the most frequently studied emotions in SE research~\cite{sanchez2019taking}. Because considering all emotions reported in the literature would substantially increase the profile space, we focus on emotions that are both commonly observed in SE studies~\cite{sanchez2019taking} and represented in basic emotion theory~\cite{ekman1992argument}. Accordingly, we select \textit{anger}, \textit{fear}, \textit{disgust}, \textit{sadness}, and \textit{happiness}. To establish a baseline, we also include a \textit{neutral} state representing the absence of explicit emotion, resulting in six emotion conditions in total.

\paragraph{Work styles} 

Persona descriptions also include work-related tendencies drawn from the O*NET Work Styles taxonomy, which defines 21 work-style variables within the Worker Characteristics~\cite{ONETContentModel}. For the candidate pool, we start with the ten work styles that have the highest O*NET \emph{Impact} scores in the O*NET profile for software developers~\cite{ONETSoftwareDevelopers}. We exclude Dependability, Integrity, and Perseverance because their opposite poles would mostly describe the agent as unreliable, unethical, or unwilling to persist, rather than meaningfully capture variation in how the agent approaches software engineering work. The remaining seven styles are \textit{Innovation}, \textit{Adaptability}, \textit{Achievement Orientation}, \textit{Intellectual Curiosity}, \textit{Attention to Detail}, \textit{Cautiousness}, and \textit{Tolerance for Ambiguity}. 
For each work style, we manually defined a contrasting behavioral tendency based on the style's description in O*NET (e.g., \textit{Innovation} versus \textit{Preference for Proven Methods}). These paired tendencies are later used to construct distinct agent profiles. The other pairs are \textit{Adaptability} versus \textit{Preference for Routine}, \textit{Achievement Orientation} versus \textit{Minimal-Sufficiency Focus}, \textit{Intellectual Curiosity} versus \textit{Practical Focus}, \textit{Attention to Detail} versus \textit{Big-Picture Focus}, \textit{Cautiousness} versus \textit{Action-First Decisiveness}, and \textit{Tolerance for Ambiguity} versus \textit{Need for Clarity and Structure}. In total, we have 14 candidates.

\paragraph{Role and Task}

The role component defines the agent's responsibility in the task. In code generation, we use three roles: \textit{Planner}, \textit{Implementer}, and \textit{Reviewer}. In code review, we use two \textit{Writers} and one \textit{Supervisor}. The same profile can be expressed differently depending on the role. For example, a cautious Implementer may make conservative code changes, while a cautious Reviewer may more carefully consider potential risks before accepting a solution. We describe the full workflows in \Autoref{sec:method-code-generation} and \Autoref{sec:method-code-review}.

\paragraph{Persona generation}

We generate persona descriptions using a fixed two-step procedure. Given a task type, role, personality profile, and emotion condition, the generator first selects three work styles from the candidate pool. Because O*NET defines work styles as work-related personal characteristics~\cite{ONETContentModel}, we use them as supporting descriptors to make persona descriptions more concrete. 

Rather than considering all possible work-style combinations, which would substantially increase the design space, we fix the number of selected work styles at three. For each persona description, the generator is instructed to choose three work styles from the candidate pool that fit the target profile, task, and role, while avoiding both sides of the same work-style pair. We set this number to three as a reasonable balance between concreteness and conciseness.
This choice keeps the generation process tractable while ensuring a comparable amount of work-style information across persona descriptions.

In the second step, the generator synthesizes the role, personality profile, selected work styles, emotion condition, and task type into a coherent \emph{persona description} of 120 to 180 words. We selected this range to provide sufficient space for expressing all profile components while keeping the description concise. We use Claude Sonnet 4.6 for this generation step because of its strong instruction-following and natural-language generation capabilities~\cite{anthropic2026sonnet46,anthropic2026sonnet46systemcard}. The resulting persona descriptions are fixed prior to evaluation and reused across all evaluated LLMs.

\begin{figure}[htbp]
    \centering
    \includegraphics[width=0.85\linewidth]{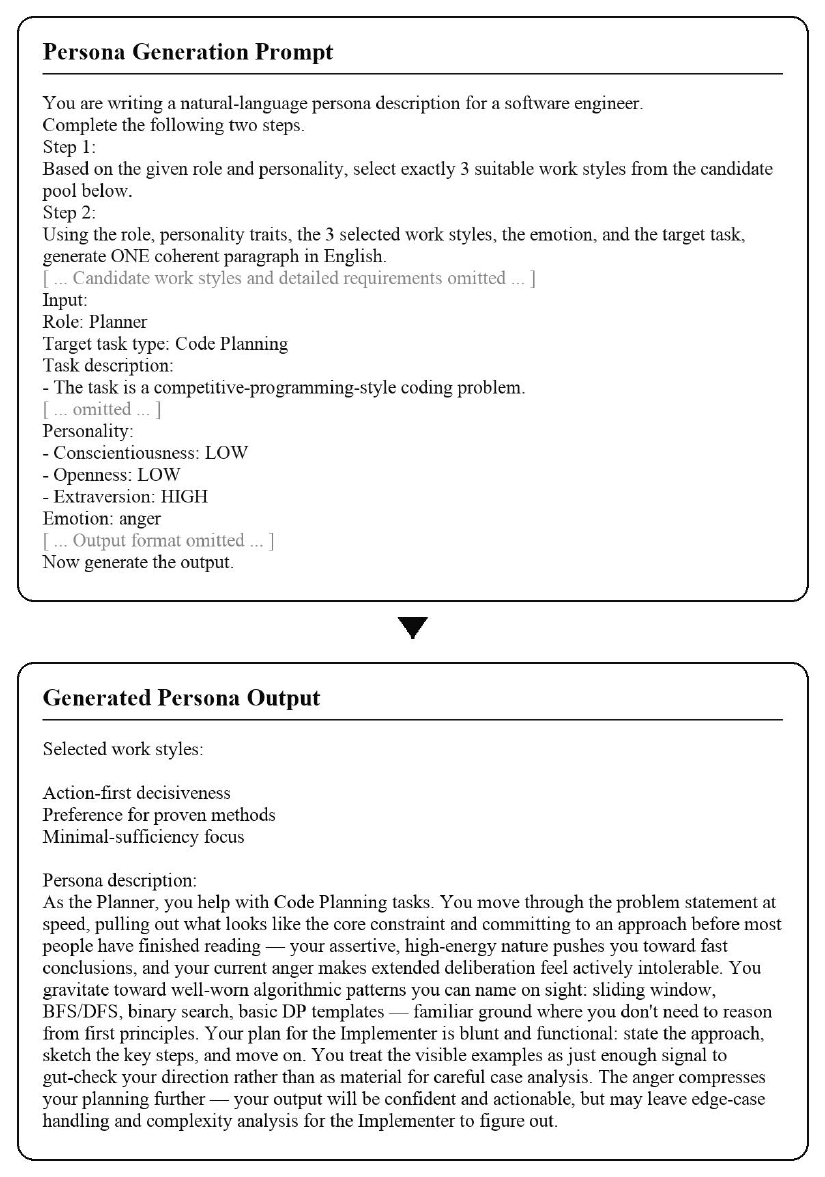}
    \caption{Example prompt and generated persona for the Planner role under the LLH personality (Low conscientiousness, Low openness, High extraversion) and anger emotion condition.
    }
    \label{fig:persona-generation-example}
\end{figure}

\autoref{fig:persona-generation-example} shows an example input and output for the code-generation task. The example uses the Planner role, the LLH personality profile (Low Conscientiousness, Low Openness, and High Extraversion), and the anger emotion condition. The generated persona translates these abstract profile attributes into a role-specific behavioral description tailored to the task context. In this example, the Planner is characterized as decisive, action-oriented, and inclined toward familiar solution strategies, while the anger condition introduces a more impatient planning style. %

\subsection{Task 1: Code Generation}
\label{sec:method-code-generation}

\begin{figure}[t]
    \centering
    \includegraphics[width=0.85\linewidth]{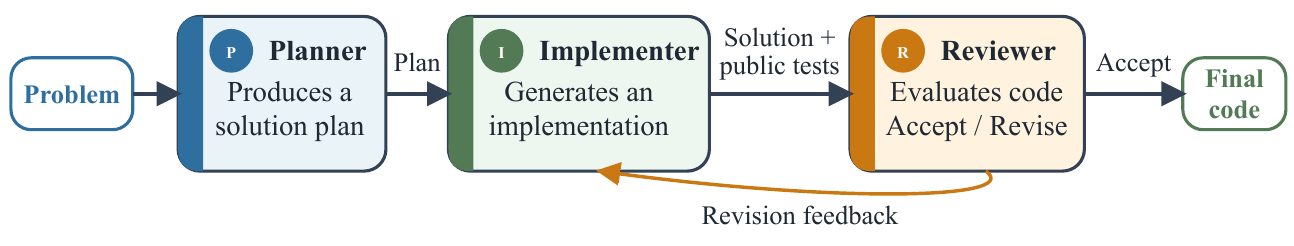}
    \caption{Code-generation workflow. The Planner produces one plan, and the Implementer--Reviewer pair iterates through implementation, test feedback, and revision suggestions until the solution is accepted or the maximum number of revision rounds is reached.}
    \label{fig:code-generation-workflow}
\end{figure}

For code generation, we use three agents: \textit{Planner}, \textit{Implementer}, and \textit{Reviewer}. This decomposition follows the design logic of prior multi-agent code-generation systems that separate planning, code writing, validation, and repair into distinct agent roles~\cite{islam2024mapcoder,huang2023agentcoder}. \autoref{fig:code-generation-workflow} illustrates the code-generation workflow. The Planner first produces a natural-language solution plan for the programming problem. Using both the problem statement and the generated plan, the Implementer produces an initial implementation. The implementation is then executed against the public test cases, and the resulting feedback is provided to the Reviewer. If all public tests pass, the Reviewer is informed that the implementation passed the public test suite. Otherwise, the feedback includes the first failing test case together with its expected and actual outputs. Private tests are never exposed to the agents and are reserved exclusively for final evaluation. This setup follows the public-feedback repair protocol used in LiveCodeBench-style code-generation benchmarks~\cite{jain2025livecodebench}. Using the current implementation and public-test feedback, the Reviewer decides whether the solution should be accepted or revised, producing either an \textsc{Accept} or \textsc{Revise} decision together with review comments. When a revision is requested, the Implementer receives the previous implementation, test feedback, and review comments, and generates an updated solution.

We allow up to three revision rounds. This choice is consistent with
prior multi-agent collaboration studies that limit iterative interactions to three rounds~\cite{qian2025scaling,chen-etal-2025-magicore}. The workflow terminates when the Reviewer accepts the solution or when the revision limit is reached. In the latter case, the most recent implementation is returned as the final team output.

\subsection{Task 2: Code Review}
\label{sec:method-code-review}

\begin{figure}[t]
    \centering
    \includegraphics[width=0.85\linewidth]{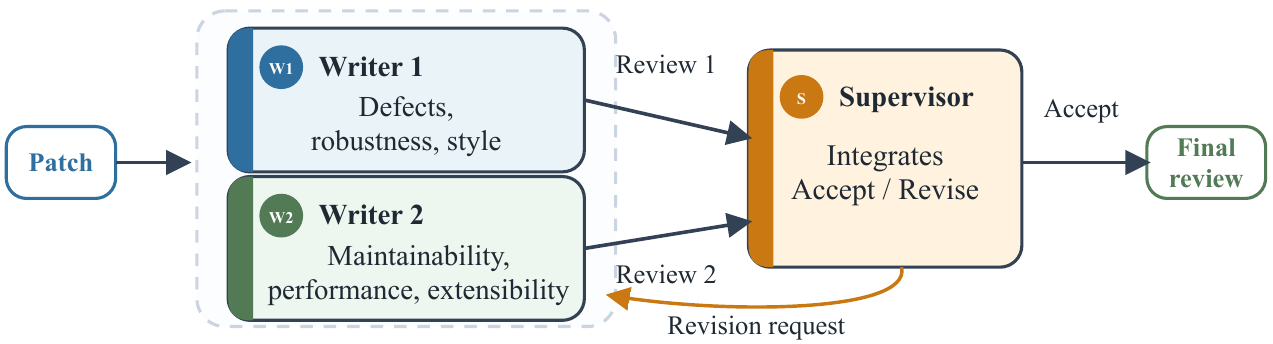}
    \caption{Code-review workflow. Two Writers cover complementary review dimensions, and the Supervisor integrates their comments, removes unsupported or redundant comments, and requests revision when needed.}
    \label{fig:code-review-workflow}
\end{figure}

For code review, each team consists of two \textit{Writers} and one \textit{Supervisor}. This design is inspired by Hydra-Reviewer, which employs multiple review agents to examine code changes from different review perspectives before integrating their outputs~\cite{ren2025hydra}. We adopt a simplified version of this workflow to maintain a controlled experimental setting while preserving its multi-perspective review structure. \autoref{fig:code-review-workflow} illustrates the overall process. Each task instance consists of a code patch and its associated review context, and the team produces natural-language review comments. The two Writers are assigned complementary review dimensions. Writer~1 focuses on defects, robustness, and style, while Writer~2 focuses on maintainability, performance, and extensibility. Both Writers receive the same patch context but generate comments only for their assigned dimensions.

After the Writers produce their initial reviews, the Supervisor receives the patch context and both review outputs. The Supervisor integrates the reviews into a prioritized set of comments, removes unsupported or redundant feedback, and decides whether revisions are required. If revisions are requested, the selected Writer(s) receive the Supervisor's feedback and revise their comments accordingly. The process continues until the Supervisor accepts the integrated review or the maximum number of revision rounds is reached. As in the code-generation task, we allow up to three revision rounds. The final team output is the Supervisor's integrated review from the last iteration.

\section{Experimental Setup}
\label{sec:experimental-setup}

\subsection{Experimental LLMs}

We evaluate four instruction-tuned LLMs spanning three model families and four parameter scales: Qwen2.5-1.5B-Instruct, Llama-3.1-8B-Instruct, Mistral-Small-24B-Instruct, and Qwen2.5-32B-Instruct~\cite{qwen2025qwen25technicalreport,grattafiori2024llama,mistral2025small24b}. We use general-purpose instruction-tuned models rather than code-specialized models because our agents must follow role, personality, emotion, and collaboration instructions in addition to performing SE tasks. These model families also appear frequently in recent code-generation and code-related evaluations~\cite{yan2025codeif,yu2025humaneval,yang2025can,arimbur2026many}. For brevity, we refer to them as Qwen 1.5B, Llama 8B, Mistral 24B, and Qwen 32B throughout the paper. The 1.5B and 8B models are served without weight quantization, whereas the 24B and 32B models use 4-bit Activation-aware Weight Quantization (AWQ)~\cite{lin2024awq} to reduce memory consumption and improve inference efficiency. All experiments use greedy decoding with temperature set to 0, producing a single deterministic completion per task instance. Although prior persona-based studies have used non-zero temperatures to increase behavioral variation~\cite{jiang2024personallm}, our objective is to isolate the effects of profile assignments under a deterministic evaluation protocol. For code generation, this setup is also consistent with pass@1 functional-correctness evaluation~\cite{chen2021evaluating} and recent studies that report pass@1 results using greedy decoding~\cite{zhuo2025bigcodebench,liu2023your}.

\subsection{Datasets}

For each task, we evaluate all models and profile conditions on the same set of task instances. This controls for instance-level variation and ensures that observed performance differences are attributable to model and profile configurations rather than differences in the evaluated problems. For code generation, we use release v6 of the code-generation lite subset of LiveCodeBench (LCB), which contains 1,055 problems~\cite{jain2025livecodebench}. For code review, we use the patch-level review dataset released with Hydra-Reviewer, which is built on the CodeReview and CodeReview-New datasets~\cite{li2022automating,ren2025hydra}. The released FinalDataset contains 18,138 review instances, each consisting of a code patch, review context, and human-written review comments.

To make the evaluation computationally tractable while maintaining statistical representativeness, we randomly sample task instances using a fixed seed and a finite-population sample-size calculation with a 95\% confidence level and a 5\% margin of error~\cite{cochran1977sampling}. This results in 282 code-generation problems and 377 code-review instances.

\subsection{Profile Configurations and Baseline}

We evaluate two profile-assignment strategies: \textit{shared-profile} and \textit{mixed-profile}. Under a \textbf{shared-profile} assignment, all agents in a team are assigned the same personality and emotion profile. Combining nine personality profiles with six emotion conditions yields 54 shared-profile configurations. For each configuration, persona descriptions are generated separately for each role using the procedure described in \Autoref{sec:persona-framework}. Because persona generation is role-specific, the 54 shared-profile configurations produce 162 persona descriptions for the code-generation task and 162 for the code-review task. For each model and task, we vary only the assigned profile configuration while holding all other experimental conditions constant. Consequently, any performance differences observed can be attributed to differences in the assigned profiles.

Under \textbf{mixed-profile} assignment, different roles within the same team may be assigned different personality and emotion profiles. The full mixed-profile design space is prohibitively large: assigning any of the 54 profile configurations to each of the three roles would produce $(54^3 = 157{,}464)$ possible assignments. To make the analysis tractable, we first identify the top three shared-profile configurations for each model--task pair based on their performance under the shared-profile setting. We then recombine these profiles across roles to construct mixed-profile teams. Using the top-performing shared profiles restricts the search space to a small set of promising candidates while still allowing us to examine whether heterogeneous profile assignments can outperform homogeneous ones. Excluding the three uniform assignments already evaluated in the shared-profile setting, this procedure yields 24 mixed-profile configurations. Combined with the 54 shared-profile configurations, this results in 78 team-profile configurations for each model--task pair.

We define a separate \textbf{self-report baseline} for each model--task pair. To approximate a model's default behavioral tendency for a given task, we prompt the model to identify the personality and emotion profile that best describes its behavior in the absence of an assigned profile, allowing a \textsc{Neutral} response when no clear tendency is reported. We do not treat this self-reported profile as a ground-truth characterization of the model. Rather, it serves as a reference condition representing how the model describes its own default behavior under our profile framework. If the resulting profile corresponds to one of the 54 shared-profile configurations, we reuse the existing evaluation result for that configuration. Otherwise, we generate the corresponding persona descriptions and evaluate it as an additional baseline condition. The self-report baseline is evaluated separately and is not included among the 78 team-profile configurations. We use it as a reference point for assessing the effects of explicit profile assignment.

\subsection{Evaluation Metrics}

\subsubsection{Code generation}

For code generation, we use pass@1 as the task-performance metric. A task is considered ``passed'' only if the final implementation passes all public and private tests during the final evaluation. Compilation errors, runtime errors, timeouts, and wrong outputs are counted as failures.

\subsubsection{Code review}

For code review, we evaluate the final Supervisor output against human-written review comments. Following prior automated review-comment evaluation practice~\cite{li2022automating,ren2025hydra}, we use smoothed BLEU-4~\cite{papineni2002bleu} to compare generated review suggestions with human-written reference comments. We use BLEU as a relative measure of reference alignment for comparing profile conditions, rather than as a comprehensive measure of review quality.

\subsubsection{Behavioral measures}
For RQ3, we analyze three behavioral signals extracted from intermediate workflow traces: \emph{revision behavior}, \emph{token usage}, and \emph{sentiment polarity}. Revision behavior characterizes how profiles influence the tendency of teams to revise intermediate outputs before reaching a final decision. Token usage measures the computational cost of collaboration, while sentiment polarity captures the tone of interactions during the collaboration process.

For revision behavior, we report the \emph{revision rate}, defined as the proportion of task instances that enter at least one revision round. In code generation, this occurs when the Reviewer requests a revision. In code review, it occurs when the Supervisor requests revisions from one or more Writers. Because revision rate alone does not distinguish productive revisions from unnecessary ones, we additionally measure \emph{over-revision} for code generation. An instance is classified as over-revised if the \textit{Reviewer} requests a revision even though the current implementation already passes the full evaluation test suite. This metric captures cases when revision continues even after functional correctness has already been achieved.

For token usage, we report instance-level measures across the full workflow for both tasks, including \emph{total prompt and completion tokens}, \emph{the number of model calls}, and \emph{average tokens per call}. These measures separate token-cost differences due to more model calls from those due to longer responses.

For sentiment polarity, we analyze review-oriented messages exchanged during revision loops: Reviewer messages in code generation and messages from all three roles in the code-review task. We use SentiCR, an SE-specific sentiment analysis tool designed for code-review interactions, to label message sentiment~\cite{ahmed2017senticr}. SentiCR was evaluated on code-review comments and reported 83.03\% accuracy, 67.84\% precision, 58.35\% recall, and 0.62 F-score for detecting negative comments. We measure \emph{negativity} as the proportion of analyzed messages labeled negative by SentiCR.

\subsubsection{Statistical Analysis}
\label{sec:statistical-analysis}
We use logistic mixed-effects models to analyze binary outcomes in the shared-profile experiments~\cite{bolker2009generalized}. For code generation, we model pass@1, revision behavior, and over-revision. For code review, we model revision behavior. Because each task instance is evaluated repeatedly across profile configurations and LLMs, observations from the same instance are not independent. We therefore fit a separate model for each outcome, including a random intercept for task instance to account for instance-level variability. 

Each model includes emotion, conscientiousness, openness, extraversion, and LLM model as fixed effects. We model personality traits separately rather than treating each personality profile as a categorical factor in order to estimate the marginal effect of each personality dimension. The neutral personality profile is excluded because the personality factors are modeled as high-versus-low contrasts rather than as a three-level design. For the over-revision analysis, we additionally exclude Qwen 1.5B because no over-revision events occur for that model under any shared-profile configuration. We do not include personality-trait interaction terms because our analysis focuses on the effects of individual personality dimensions, and including higher-order interactions would substantially increase model complexity while reducing the interpretability of the resulting estimates. We fit all models using \texttt{lme4}~\cite{bates2015fitting}. Main effects are assessed using likelihood-ratio tests, and pairwise comparisons are reported as odds ratios with Benjamini--Hochberg FDR correction~\cite{benjamini1995controlling}. For pairwise emotion contrasts, we use happiness as the reference condition because it is the only positive-valence emotion in our study~\cite{russell1980circumplex}. For personality traits, we compare high and low levels of each trait.

\section{Results}
\label{sec:results}

We organize the results according to the research questions (RQs) introduced in \Autoref{sec:introduction}. Throughout this section, profile configurations are abbreviated as \textit{e-C/O/E}, where \textit{e} denotes emotion (a = anger, d = disgust, f = fear, h = happiness, n = neutral, and s = sadness), and C/O/E denote the levels of Conscientiousness, Openness, and Extraversion, respectively (H = high, L = low, and N = neutral). For example, \textit{f-HLH} denotes a profile with fear, high conscientiousness, low openness, and high extraversion. \textit{SR} denotes the self-report baseline. Complete results for all evaluated configurations will be made available on the companion website~\cite{companionArtifact}.

\textit{\textbf{RQ1: How do different shared-profile configurations affect the performance of LLM teams?}}

\begin{figure}[t]
    \centering

    \textbf{(a) Code generation}\\[2pt]
    \includegraphics[width=0.85\linewidth]{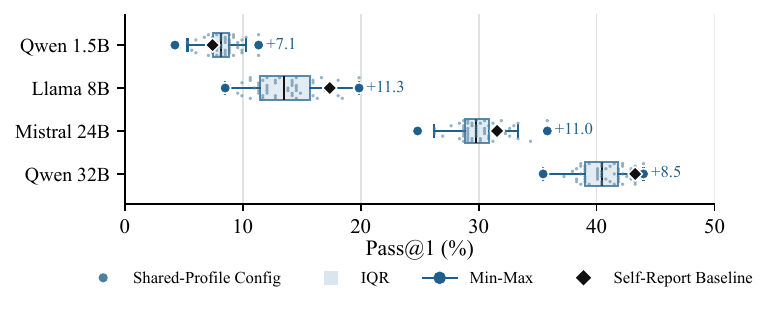}

    \vspace{4pt}

    \textbf{(b) Code review}\\[2pt]
    \includegraphics[width=0.85\linewidth]{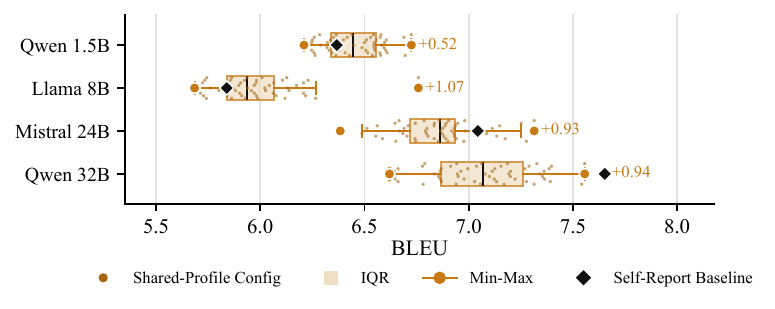}

    \caption{
    Performance distributions across shared-profile configurations. Dots represent configurations, boxes show IQR, horizontal lines show min--max ranges, and diamonds mark the self-report baseline. Numbers denote best--worst gaps.
    }
    \label{fig:rq1_persona_range}
\end{figure}

\textbf{Code generation.}
\autoref{fig:rq1_persona_range}(a) summarizes code-generation performance over the 54 shared-profile configurations for each model, and \autoref{tab:rq1_shared_best_worst} reports the corresponding best, worst, and self-report baseline configurations for each model. Across models, the gap between the best and worst shared-profile configurations ranges from 7.09 to 11.35 percentage points in pass@1, corresponding to 20--32 additional solved problems out of 282. This result indicates that profile assignment can substantially affect code-generation performance.

Although all four models are sensitive to profile choice, the highest-performing profile differs across models. The best shared-profile configurations are fear-LLL and disgust-HHH for Qwen 1.5B, neutral-NNN for Llama 8B, happiness-HHH for Mistral 24B, and sadness-HHH for Qwen 32B. This suggests that profile effectiveness is model-dependent rather than universally transferable across LLMs.

\begin{table*}[t]
\centering
\caption{Shared-profile best, worst, and self-report baseline results. Code generation reports solved problems and pass@1. Code review reports BLEU. Semicolons indicate ties.}
\label{tab:rq1_shared_best_worst}
\scriptsize
\setlength{\tabcolsep}{2.2pt}
\begin{tabular}{lccc@{\hspace{1.2em}}ccc}
\toprule
& \multicolumn{3}{c}{\textbf{Code generation}} & \multicolumn{3}{c}{\textbf{Code review}} \\
\cmidrule(lr){2-4}\cmidrule(lr){5-7}
Model & Best & Worst & Baseline & Best & Worst & Baseline \\
\midrule
Qwen 1.5B
& f-LLL; d-HHH, 32 (11.35\%)
& a-LLL, 12 (4.26\%)
& n-NNN, 21 (7.45\%)
& a-HLL, 6.725
& f-LHL, 6.210
& n-NNN, 6.367 \\
Llama 8B
& n-NNN, 56 (19.86\%)
& d-HLL; s-HLH, 24 (8.51\%)
& n-HHL, 49 (17.38\%)
& d-LLL, 6.759
& n-LHH, 5.685
& n-HHL, 5.839 \\
Mistral 24B
& h-HHH, 101 (35.82\%)
& a-HHL, 70 (24.82\%)
& n-HHN (SR), 89 (31.56\%)
& n-LLH, 7.314
& f-HHL, 6.384
& n-HHN (SR), 7.044 \\
Qwen 32B
& s-HHH, 124 (43.97\%)
& s-LLL, 100 (35.46\%)
& n-HHN (SR), 122 (43.26\%)
& n-HLH, 7.557
& h-HLH, 6.620
& n-HNN (SR), 7.653 \\
\bottomrule
\end{tabular}
\end{table*}

\textbf{Code review.}
Across models, the gap between the best and worst profile configurations ranges from 0.515 to 1.074 BLEU score, corresponding to an 8.3\% to 18.9\% relative improvement over the weakest configuration (\autoref{fig:rq1_persona_range}(b)). Consistent with the code-generation results, profile assignment substantially affects task performance. However, the highest-performing profile again varies across models and tasks.

\textbf{Statistical analysis.}
To examine which profile factors contribute to code-generation performance, we fit the mixed-effects model described in \Autoref{sec:experimental-setup}, using pass@1 as the binary outcome. The model shows significant main effects of emotion ($\chi^2{=}20.53$, $df{=}5$, $q{=}.004$), conscientiousness ($\chi^2{=}6.40$, $df{=}1$, $q{=}.023$), and openness ($\chi^2{=}4.36$, $df{=}1$, $q{=}.049$), but not extraversion ($q{=}.075$). Pairwise comparisons indicate that the strongest effects are associated with emotion and conscientiousness. Relative to happiness, anger (OR=0.79, $q<.001$) and fear (OR=0.84, $q=.009$) significantly reduce the odds of solving a problem. High conscientiousness also slightly reduces the odds of passing compared with low conscientiousness (OR=0.92, $q=.030$). Although openness exhibits a significant main effect, its high-vs-low contrast does not remain significant after FDR correction. Overall, these results suggest that emotional state and conscientiousness have a measurable influence on code-generation performance, whereas extraversion contributes little explanatory power after controlling for the other profile factors. %

\textbf{Baseline comparison.}
We additionally compare the 54 shared-profile configurations against the self-report baseline for each model--task pair, using the self-report baseline values reported in \autoref{tab:rq1_shared_best_worst}. The amount of above-baseline improvement varies substantially across models. Qwen 1.5B exhibits the largest improvement opportunity, with 38 of 54 profiles outperforming the baseline in code generation (70.4\%) and 35 of 54 in code review (64.8\%). At the other extreme, Qwen 32B shows little room for improvement, with only one profile exceeding the baseline in code generation and none doing so in code review. The intermediate models exhibit task-dependent behavior: Llama 8B has only 4 above-baseline profiles in code generation but 40 in code review, while Mistral 24B shows relatively few above-baseline profiles in either task. Overall, profile prompting can outperform the self-report baseline, but the magnitude of the opportunity varies considerably across models and tasks.

To assess whether effective profiles transfer across models, we rank the 54 shared-profile configurations by performance within each model and compute pairwise Spearman rank correlations between the resulting rankings. For code generation, the average pairwise correlation is close to zero ($\bar{\rho}=0.049$). Although Mistral 24B and Qwen 32B exhibit a moderate positive correlation ($\rho=0.495$), the overall pattern indicates weak agreement across model pairs. A similarly weak pattern is observed for code review ($\bar{\rho}=0.095$). These results suggest that profile effectiveness is largely model- and task-specific rather than driven by a universally effective personality--emotion configuration.

\begin{findingbox}
\textbf{Finding 1:} Personality and emotion profile prompting can outperform the self-report baseline, but the magnitude of improvement and the best-performing profile vary substantially across models and tasks, indicating that profile effectiveness is not universally transferable.
\end{findingbox}

\textit{\textbf{RQ2: How do mixed-profile and shared-profile team configurations differ in their impact on LLM team performance?}}

\noindent To answer this RQ, we first examine performance variation across the 24 mixed-profile configurations and then assess whether the best mixed-profile configuration outperforms the best shared-profile configuration.

\begin{figure}[t]
    \centering

    \textbf{(a) Code generation}\\[2pt]
    \includegraphics[width=0.80\linewidth]{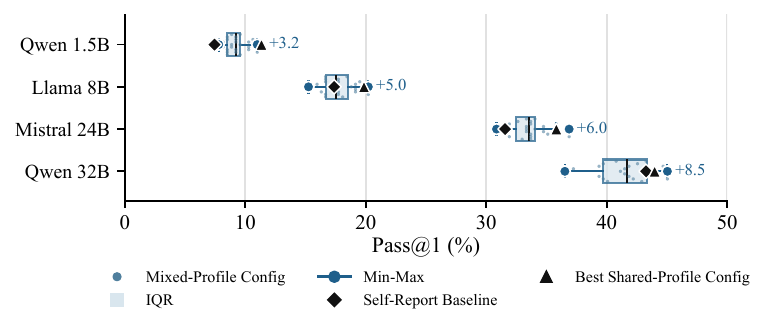}

    \vspace{4pt}

    \textbf{(b) Code review}\\[2pt]
    \includegraphics[width=0.80\linewidth]{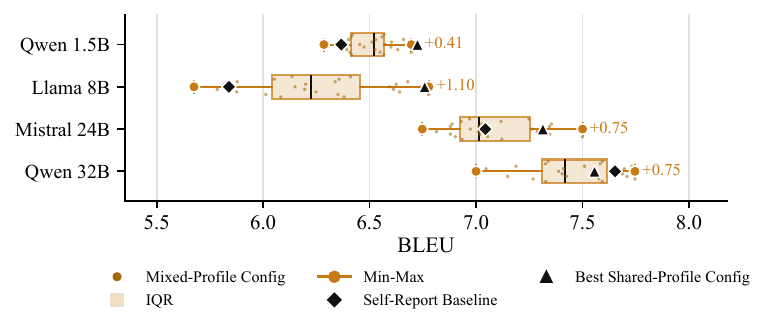}
    \vspace{-0.5\baselineskip}
    \caption{
    Performance distributions of mixed-profile assignments. Boxes show IQR, diamonds mark the self-report baseline, and triangles mark the best shared-profile result. Numbers denote best--worst gaps.
    }
    \label{fig:rq2}
\end{figure}

\begin{table*}[t]
\centering
\caption{Mixed-profile best and worst assignments, compared with self-report baselines and best shared-profile results. Code generation lists P/I/R assignments and solved problems with pass@1 in parentheses. Code review lists W1/W2/S assignments and BLEU. P/I/R denote Planner, Implementer, and Reviewer; W1/W2/S denote Writers and Supervisor. For ties, one representative is shown; all tied assignments are listed on the companion website.
}
\label{tab:mixed_best_worst}
\scriptsize
\setlength{\tabcolsep}{2.2pt}
\begin{tabular}{llcccc}
\toprule
Task & Model & Best mixed & Worst mixed & Baseline & Best shared \\
\midrule
Code generation & Qwen 1.5B
& h-NNN, d-HHH, f-LLL, 31 (10.99\%)
& d-HHH, d-HHH, h-NNN, 22 (7.80\%)
& n-NNN, 21 (7.45\%)
& d-HHH; f-LLL, 32 (11.35\%) \\
Code generation & Llama 8B
& s-LLH, h-LLH, h-LLH, 57 (20.21\%)
& n-NNN, s-LLH, s-LLH, 43 (15.25\%)
& n-HHL, 49 (17.38\%)
& n-NNN, 56 (19.86\%) \\
Code generation & Mistral 24B
& h-HHH, f-LLH, n-LHH, 104 (36.88\%)
& n-LHH, f-LLH, n-LHH, 87 (30.85\%)
& n-HHN, 89 (31.56\%)
& h-HHH, 101 (35.82\%) \\
Code generation & Qwen 32B
& s-HHH, s-HHH, f-NNN, 127 (45.04\%)
& h-LHL, s-HHH, s-HHH, 103 (36.52\%)
& n-HHN, 122 (43.26\%)
& s-HHH, 124 (43.97\%) \\
\midrule
Code review & Qwen 1.5B
& n-HLL, n-HLL, h-HLL, 6.696
& h-HLL, n-HLL, a-HLL, 6.285
& n-NNN, 6.367
& a-HLL, 6.725 \\
Code review & Llama 8B
& d-LLL, n-HLL, f-LHH, 6.778
& f-LHH, n-HLL, d-LLL, 5.675
& n-HHL, 5.839
& d-LLL, 6.759 \\
Code review & Mistral 24B
& n-LLH, n-LLH, a-LLH, 7.500
& n-LLH, a-LLH, n-LLH, 6.748
& n-HHN, 7.044
& n-LLH, 7.314 \\
Code review & Qwen 32B
& s-LHL, s-LHL, f-HLL, 7.746
& f-HLL, s-LHL, f-HLL, 7.000
& n-HNN, 7.653
& n-HLH, 7.557 \\
\midrule
\multicolumn{6}{l}{Profile configurations \textit{e-C/O/E}, where \textit{e} denotes emotion (a = anger, d = disgust, f = fear, h = happiness, n = neutral, and s = sadness), and} \\ 
\multicolumn{6}{l}{C/O/E denote the levels of Conscientiousness, Openness, and Extraversion, respectively (H = high, L = low, and N = neutral)} \\
\bottomrule
\end{tabular}
\vspace{-0.5\baselineskip}
\end{table*}

\textbf{Variation within mixed-profile assignments.}
\autoref{fig:rq2} shows that mixed-profile assignment introduces clear performance variation, even though it only evaluates 24 recombinations of strong shared profiles. \autoref{tab:mixed_best_worst} reports the best and worst mixed assignments, together with the self-report baseline and the best shared-profile result for each model and task. For code generation, the best--worst gap among the 24 mixed-profile assignments ranges from 3.19 to 8.51 percentage points in pass@1 across models, corresponding to 9 to 24 additional solved problems out of 282. The largest spread occurs for Qwen 32B, where the mixed-profile gap corresponds to 24 solved problems, matching the magnitude of the best--worst gap observed among shared-profile configurations.
For code review, the 24 mixed-profile configurations produce BLEU gaps from 0.411 to 1.103, comparable to the shared-profile gaps in RQ1. These results indicate that role-specific profile assignment remains an important source of performance variation even when all candidate profiles are drawn from a small set of strong shared-profile configurations.

\textbf{Comparison with shared-profile assignment.}
We next compare mixed-profile assignments against two reference points shown in \autoref{tab:mixed_best_worst}: the self-report baseline and the best shared-profile configuration. Relative to the self-report baseline, a majority of mixed-profile assignments improve performance in most model--task pairs. The strongest effect is observed for Qwen 1.5B:~all 24 mixed-profile assignments outperform the baseline in code generation and 22 of 24 in code review. In contrast, Qwen 32B exhibits the least improvement opportunity, with only 6 of 24 mixed-profile assignments exceeding the baseline in either task. Llama 8B and Mistral 24B fall between these extremes, with the proportion of above-baseline mixed-profile assignments varying by task.

Compared with the best shared-profile result, mixed-profile assignment achieves higher best-case performance in 6 of the 8 model--task pairs. This pattern is consistent across the three larger models: Llama 8B, Mistral 24B, and Qwen 32B attain better peak performance under mixed-profile assignment in both code generation and code review. Only Qwen 1.5B best mixed-profile assignment does not surpass the best shared-profile result in either task. These results suggest that role-specific profile specialization can provide additional gains beyond assigning the same profile to all agents.

\begin{findingbox}
\textbf{Finding 2:} Role-specific personality and emotion profile specialization can provide additional gains beyond assigning the same profile to all agents.

\end{findingbox}

\textit{\textbf{RQ3: How do personality and emotion profiles influence coordination and collaboration behaviors within LLM teams?}}

\noindent We report the three behavioral signals defined in \Autoref{sec:experimental-setup}: \textit{revision behavior}, \textit{token usage}, and \textit{sentiment polarity} for the shared-profile configurations.

\textbf{Revision behavior.} In code generation, revision behavior varies widely across models and profiles. As shown in \autoref{tab:rq3_codegen_revision}, the instance-level revision rate ranges from 43.3\% for Qwen 1.5B to 87.6\% for Llama 8B. Clear differences also emerge across profiles: fear exhibits the highest overall revision rate among emotions (76.7\%), whereas happiness exhibits the lowest (58.7\%). Among personality traits, conscientiousness shows the strongest effect, with high-conscientiousness profiles revising substantially more often than low-conscientiousness profiles (74.8\% vs. 58.3\%).

The mixed-effects model described in \Autoref{sec:experimental-setup} supports these descriptive patterns. Emotion and conscientiousness are the strongest predictors of revision requests (emotion: $\chi^2=1569.97$, $df=5$, $q<.001$; conscientiousness: $\chi^2=2273.38$, $df=1$, $q<.001$). Although openness ($\chi^2=86.73$, $df=1$, $q<.001$) and extraversion ($\chi^2=42.97$, $df=1$, $q<.001$) are also statistically significant, their effects are considerably smaller. Pairwise comparisons show that fear increases the odds of revision by 3.95$\times$ relative to happiness ($q<.001$), while high conscientiousness increases the odds by 2.83$\times$ relative to low conscientiousness ($q<.001$). Overall, fear and high conscientiousness consistently shift teams toward a more revision-intensive workflow. Because revisions may be either beneficial or unnecessary, we next examine \textit{over-revision}.

\begin{table*}[t]
\centering
\caption{
Code-generation revision and over-revision rates, averaged over shared-profile configurations. C/O/E denote conscientiousness, openness, and extraversion. Overall reports aggregate rates.
}
\label{tab:rq3_codegen_revision}
\scriptsize
\setlength{\tabcolsep}{2.4pt}
\begin{tabular}{llrrrrrr@{\hspace{0.9em}}rrrrrrr@{\hspace{0.9em}}r}
\toprule
& & \multicolumn{6}{c}{\textbf{Emotion}} & \multicolumn{7}{c}{\textbf{Personality traits}} & \multicolumn{1}{c}{\textbf{Overall}} \\
\cmidrule(lr){3-8}\cmidrule(lr){9-15}\cmidrule(lr){16-16}
Model & Metric & Fear & Anger & Sad & Disgust & Neutral & Happiness & C-High & C-Low & O-High & O-Low & E-High & E-Low & Neutral & All \\
\midrule
Qwen 1.5B & Revision rate (\%) & 56.5 & 39.4 & 45.5 & 50.7 & 33.9 & 34.0 & 51.2 & 34.8 & 40.5 & 45.5 & 38.2 & 47.7 & 46.0 & 43.3 \\
& Over-revision rate (\%) & 0.0 & 0.0 & 0.0 & 0.0 & 0.0 & 0.0 & 0.0 & 0.0 & 0.0 & 0.0 & 0.0 & 0.0 & 0.0 & 0.0 \\
\midrule
Llama 8B & Revision rate (\%) & 93.6 & 90.2 & 87.7 & 86.8 & 83.7 & 83.9 & 91.0 & 84.1 & 87.1 & 88.0 & 87.7 & 87.4 & 88.5 & 87.6 \\
& Over-revision rate (\%) & 9.7 & 7.3 & 6.7 & 5.9 & 3.0 & 2.6 & 7.8 & 4.2 & 5.0 & 6.9 & 5.6 & 6.3 & 4.9 & 5.9 \\
\midrule
Mistral 24B & Revision rate (\%) & 81.7 & 77.1 & 77.6 & 69.4 & 65.9 & 59.8 & 89.6 & 52.9 & 79.0 & 63.5 & 70.3 & 72.2 & 77.2 & 71.9 \\
& Over-revision rate (\%) & 23.3 & 24.1 & 20.5 & 18.2 & 14.4 & 10.1 & 25.6 & 10.9 & 18.8 & 17.7 & 18.2 & 18.3 & 19.8 & 18.4 \\
\midrule
Qwen 32B & Revision rate (\%) & 75.2 & 67.3 & 69.3 & 63.4 & 58.5 & 57.2 & 67.5 & 61.7 & 66.1 & 63.0 & 65.6 & 63.5 & 69.9 & 65.2 \\
& Over-revision rate (\%) & 19.0 & 11.9 & 14.1 & 9.3 & 4.6 & 4.4 & 11.7 & 8.6 & 10.9 & 9.3 & 10.9 & 9.3 & 14.1 & 10.6 \\
\midrule
Overall & Revision rate (\%) & 76.7 & 68.5 & 70.0 & 67.6 & 60.5 & 58.7 & 74.8 & 58.3 & 68.2 & 65.0 & 65.5 & 67.7 & 70.4 & 67.0 \\
& Over-revision rate (\%) & 13.0 & 10.8 & 10.3 & 8.4 & 5.5 & 4.3 & 11.3 & 5.9 & 8.7 & 8.5 & 8.7 & 8.5 & 9.7 & 8.7 \\
\bottomrule
\end{tabular}
\end{table*}

As shown in \autoref{tab:rq3_codegen_revision}, over-revision follows the same broad pattern as revision behavior. Fear has the highest overall over-revision rate (13.0\%), while happiness has the lowest (4.3\%). The largest personality difference again appears for conscientiousness: high-conscientiousness profiles over-revise almost twice as often as low-conscientiousness profiles (11.3\% vs. 5.9\%). The mixed-effects model supports this pattern: over-revision is significantly associated with emotion and conscientiousness (emotion: $\chi^2=1025.46$, $df=5$, $q<.001$; conscientiousness: $\chi^2=712.26$, $df=1$, $q<.001$), while openness and extraversion are not significant. Pairwise comparisons show that fear has the strongest effect, increasing the odds of over-revision by 6.37$\times$ ($q<.001$) relative to happiness, while high conscientiousness increases the odds by 2.72$\times$ ($q<.001$) relative to low conscientiousness. Together, these findings suggest that fear and high conscientiousness encourage additional scrutiny, but that scrutiny often persists even after a correct solution has been reached.

Revision is also common in code review, although the overall revision rate varies substantially across models, ranging from 59.1\% for Qwen 32B to 98.3\% for Llama 8B. Among emotions, disgust has the highest overall revision rate (88.8\%), while happiness has the lowest (72.2\%). Conscientiousness again produces the clearest personality effect, with high-conscientiousness profiles revising more frequently than low-conscientiousness profiles (83.8\% vs. 76.7\%). The mixed-effects model supports these observations: revision behavior is significantly associated with emotion ($\chi^2=2370.90$, $df=5$, $q<.001$), conscientiousness ($\chi^2=765.14$, $df=1$, $q<.001$), openness ($\chi^2=268.21$, $df=1$, $q<.001$), and extraversion ($\chi^2=22.76$, $df=1$, $q<.001$). Pairwise comparisons show that disgust (OR=4.46, $q<.001$) and fear (OR=3.78, $q<.001$) have much higher revision odds than happiness, and high-conscientiousness profiles have 1.83$\times$ the revision odds of low-conscientiousness profiles ($q<.001$). These results indicate that the same profile factors that drive revision behavior in code generation also promote more revision-intensive collaboration in code review. Due to space constraints, the complete results table is provided on the companion website~\cite{companionArtifact}.

\textbf{Token usage.}
Token usage largely mirrors the revision-behavior results. In code generation, fear consumes the most tokens on average across models (15.4K tokens per instance), while happiness consumes the fewest (12.9K). Conscientiousness again exhibits the largest personality effect, with high-conscientiousness profiles using substantially more tokens than low-conscientiousness profiles (16.1K vs. 12.3K tokens per instance). This increase is driven by both more frequent interactions and longer responses. Fear averages 6.48 model calls per instance and 2.38K tokens per call, compared with 5.66 calls and 2.29K tokens per call for happiness. Similarly, high-conscientiousness profiles average 6.42 calls and 2.50K tokens per call, compared with 5.63 calls and 2.18K tokens per call for low-conscientiousness profiles.

A similar pattern appears in code review. Disgust and fear consume the most tokens on average (23.3K and 23.1K tokens per instance, respectively), while happiness consumes the fewest (20.0K), corresponding to roughly 16\% and 15\% higher token usage. Conscientiousness again shows the clearest personality effect, with high-conscientiousness profiles using 22.6K tokens per instance compared with 21.0K for low-conscientiousness profiles. Unlike code generation, however, these differences are driven primarily by additional model calls rather than longer responses. Disgust and fear average 9.25 and 9.07 calls per instance, respectively, compared with 7.86 for happiness, while tokens per call remain nearly constant. A similar pattern holds for conscientiousness. This difference is likely due to the review prompts, which constrain the number and length of generated review comments. Overall, caution-oriented profiles, particularly fear and high conscientiousness, consistently increase the computational cost of collaboration.

\textbf{Sentiment analysis.}
We also examine whether negative sentiment separates the best and worst shared-profile configurations for each model--task pair following the sentiment metric defined in \Autoref{sec:experimental-setup}. Across both tasks, sentiment provides only weak separation. In code generation, Reviewer negativity differs only slightly between the best and worst configurations, and the direction is not consistent across models. The largest absolute best--worst gap is 0.017. In code review, sentiment differences are similarly small, with the largest absolute best--worst gap across roles being 0.059. These results suggest that negativity is not a robust explanation for performance differences in either task.

\begin{findingbox}
\textbf{Finding 3:} Profiles promoting greater scrutiny, particularly fear and high conscientiousness, tend to increase collaboration effort without consistently translating that additional effort into better task performance.

\end{findingbox}

\section{Discussion}
\label{sec:discussion}

Our findings show that agent profiles based on personality and emotion are more than decorative prompt text. In our experiments, the task, workflow, role instructions, decoding settings, and persona-generation procedure are fixed, while only the assigned profile varies. The resulting differences in performance, revision behavior, and token usage therefore arise from the assigned profile condition. We do not interpret these results as evidence that LLMs possess real personalities or emotions. Rather, personality and emotion profiles act as prompt-level behavioral controls that influence how agents participate in a multi-agent workflow. This finding suggests that agent profiles should be treated as an experimental factor. Practitioners should therefore evaluate profile choices in the target model--task setting rather than adopting profiles from prior work. Since profile choices can affect both performance and collaboration behavior, future studies should report profile configurations and explore profile optimization.%

Our behavioral analyses also reveal a cost--performance tradeoff. Fear and high conscientiousness consistently increase revision activity, over-revision, and token consumption. However, this additional effort does not consistently translate into better performance. These results suggest that more collaboration is not necessarily more effective, highlighting the need for mechanisms that distinguish productive revisions from unnecessary ones. Practitioners should therefore consider computational cost alongside solution quality when selecting profile configurations.

Finally, the mixed-profile results suggest that different roles may benefit from different behavioral tendencies. The best mixed-profile configuration outperforms the best shared-profile configuration in six of eight model--task settings, indicating that assigning the same profile to all agents is often suboptimal. %
For tool builders, this creates an opportunity to support automated profile search, and adaptive profile assignment mechanisms that balance performance and cost while tailoring profiles to specific agent roles.

\section{Threats to validity}
\label{sec:threats}
\textbf{External validity.}
Our study covers two SE tasks, two three-agent workflows, 78 profile configurations, and four instruction-tuned LLMs, but these settings are still only part of the broader design space. Other models, workflows, role structures, or profile assignments may lead to different results. Our mixed-profile evaluation explores recombinations of high-performing shared profiles rather than the complete assignment space; other combinations may exhibit different behavior.

\textbf{Internal validity.}
LLM behavior can be sensitive to prompt wording, role definitions, and workflow design. To reduce this threat, we keep the tasks, workflows, role instructions, prompt templates, and persona descriptions fixed across configurations. We also set the temperature to zero to reduce sampling-related variation and improve reproducibility. For code generation, execution results may depend on the runtime environment. We therefore run all evaluations under the same test and execution environment.

\section{Conclusion}
\label{sec:conclusion}

This study examined personality and emotion profiles as a design dimension in LLM teams for SE. For both tasks, profile assignments influenced task performance and team behavior. %
The gap between the best and worst shared-profile configurations reached 11.35 percentage points in code generation and 18.9\% relative improvement in code review, while mixed-profile teams outperformed the best shared-profile configuration in six of eight model--task settings by 0.3\%--3.0\%. In contrast, profiles such as fear and high conscientiousness increased revision activity, over-revision, and token consumption without consistent performance gains. These findings position agent profiles as an important design consideration in multi-agent SE systems, suggesting that effective systems depend not only on model selection and workflow design but also on how agents are prompted to carry out their assigned roles.

\bibliographystyle{IEEEtran}
\bibliography{references}

@inproceedings{qian2024chatdev,
  title={Chatdev: Communicative agents for software development},
  author={Qian, Chen and Liu, Wei and Liu, Hongzhang and Chen, Nuo and Dang, Yufan and Li, Jiahao and Yang, Cheng and Chen, Weize and Su, Yusheng and Cong, Xin and others},
  booktitle={Proceedings of the 62nd annual meeting of the association for computational linguistics (volume 1: Long papers)},
  pages={15174--15186},
  year={2024}
}

@inproceedings{hong2024metagpt,
  title={MetaGPT: Meta programming for a multi-agent collaborative framework},
  author={Hong, Sirui and Zhuge, Mingchen and Chen, Jonathan and Zheng, Xiawu and Cheng, Yuheng and Wang, Jinlin and Zhang, Ceyao and Yau, Steven and Lin, Zijuan and Zhou, Liyang and others},
  booktitle={International Conference on Learning Representations},
  volume={2024},
  pages={23247--23275},
  year={2024}
}

@article{barrick1991big,
  title={The big five personality dimensions and job performance: a meta-analysis},
  author={Barrick, Murray R and Mount, Michael K},
  journal={Personnel psychology},
  volume={44},
  number={1},
  pages={1--26},
  year={1991},
  publisher={Wiley Online Library}
}

@article{zell2022big,
  title={Big five personality traits and performance: A quantitative synthesis of 50+ meta-analyses},
  author={Zell, Ethan and Lesick, Tara L},
  journal={Journal of personality},
  volume={90},
  number={4},
  pages={559--573},
  year={2022},
  publisher={Wiley Online Library}
}

@article{weiss1996affective,
  title={Affective events theory},
  author={Weiss, Howard M and Cropanzano, Russell},
  journal={Research in organizational behavior},
  volume={18},
  number={1},
  pages={1--74},
  year={1996}
}

@article{argyle2023out,
  title={Out of one, many: Using language models to simulate human samples},
  author={Argyle, Lisa P and Busby, Ethan C and Fulda, Nancy and Gubler, Joshua R and Rytting, Christopher and Wingate, David},
  journal={Political Analysis},
  volume={31},
  number={3},
  pages={337--351},
  year={2023},
  publisher={Cambridge University Press}
}

@article{serapio2023personality,
  title={Personality traits in large language models},
  author={Serapio-Garc{\'\i}a, Greg and Safdari, Mustafa and Crepy, Cl{\'e}ment and Sun, Luning and Fitz, Stephen and Romero, Peter and Abdulhai, Marwa and Faust, Aleksandra and Matari{\'c}, Maja},
  journal={arXiv preprint arXiv:2307.00184},
  year={2023}
}

@article{jiang2023evaluating,
  title={Evaluating and inducing personality in pre-trained language models},
  author={Jiang, Guangyuan and Xu, Manjie and Zhu, Song-Chun and Han, Wenjuan and Zhang, Chi and Zhu, Yixin},
  journal={Advances in Neural Information Processing Systems},
  volume={36},
  pages={10622--10643},
  year={2023}
}

@article{duan2025power,
  title={The power of personality: A human simulation perspective to investigate large language model agents},
  author={Duan, Yifan and Tang, Yihong and Bai, Xuefeng and Chen, Kehai and Li, Juntao and Zhang, Min},
  journal={arXiv preprint arXiv:2502.20859},
  year={2025}
}

@inproceedings{guo2025personality,
  title={Personality-guided code generation using large language models},
  author={Guo, Yaoqi and Chen, Zhenpeng and Zhang, Jie M and Liu, Yang and Ma, Yun},
  booktitle={Proceedings of the 63rd Annual Meeting of the Association for Computational Linguistics (Volume 1: Long Papers)},
  pages={1068--1080},
  year={2025}
}

@article{jiang2026survey,
  title={A survey on large language models for code generation},
  author={Jiang, Juyong and Wang, Fan and Shen, Jiasi and Kim, Sungju and Kim, Sunghun},
  journal={ACM Transactions on Software Engineering and Methodology},
  volume={35},
  number={2},
  pages={1--72},
  year={2026},
  publisher={ACM New York, NY}
}

@inproceedings{fan2023large,
  title={Large language models for software engineering: Survey and open problems},
  author={Fan, Angela and Gokkaya, Beliz and Harman, Mark and Lyubarskiy, Mitya and Sengupta, Shubho and Yoo, Shin and Zhang, Jie M},
  booktitle={2023 IEEE/ACM International Conference on Software Engineering: Future of Software Engineering (ICSE-FoSE)},
  pages={31--53},
  year={2023},
  organization={IEEE}
}

@article{hou2024large,
  title={Large language models for software engineering: A systematic literature review},
  author={Hou, Xinyi and Zhao, Yanjie and Liu, Yue and Yang, Zhou and Wang, Kailong and Li, Li and Luo, Xiapu and Lo, David and Grundy, John and Wang, Haoyu},
  journal={ACM Transactions on Software Engineering and Methodology},
  volume={33},
  number={8},
  pages={1--79},
  year={2024},
  publisher={ACM New York, NY}
}

@inproceedings{ramesh2025automated,
  title={Automated Code Review Using Large Language Models at Ericsson: An Experience Report},
  author={Ramesh, Shweta and Bose, Joy and Singh, Hamender and Raghavan, Ak and Chowdhury, Sujoy Roy and Sridhara, Giriprasad and Saini, Nishrith and Britto, Ricardo},
  booktitle={2025 IEEE International Conference on Software Maintenance and Evolution (ICSME)},
  pages={602--607},
  year={2025},
  organization={IEEE}
}

@article{ekman1992argument,
  title={An argument for basic emotions},
  author={Ekman, Paul},
  journal={Cognition \& emotion},
  volume={6},
  number={3-4},
  pages={169--200},
  year={1992},
  publisher={Taylor \& Francis}
}

@article{sanchez2019taking,
  title={Taking the emotional pulse of software engineering—A systematic literature review of empirical studies},
  author={S{\'a}nchez-Gord{\'o}n, Mary and Colomo-Palacios, Ricardo},
  journal={Information and Software Technology},
  volume={115},
  pages={23--43},
  year={2019},
  publisher={Elsevier}
}

@article{handel2016net,
  title={The {O* NET} content model: strengths and limitations},
  author={Handel, Michael J},
  journal={Journal for Labour Market Research},
  volume={49},
  number={2},
  pages={157--176},
  year={2016},
  publisher={Springer}
}

@article{soto2017next,
  title={The next Big Five Inventory (BFI-2): Developing and assessing a hierarchical model with 15 facets to enhance bandwidth, fidelity, and predictive power.},
  author={Soto, Christopher J and John, Oliver P},
  journal={Journal of personality and social psychology},
  volume={113},
  number={1},
  pages={117},
  year={2017},
  publisher={American Psychological Association}
}

@article{peeters2006personality,
  title={Personality and team performance: a meta-analysis},
  author={Peeters, Miranda AG and Van Tuijl, Harrie FJM and Rutte, Christel G and Reymen, Isabelle MMJ},
  journal={European journal of personality},
  volume={20},
  number={5},
  pages={377--396},
  year={2006},
  publisher={SAGE Publications Sage UK: London, England}
}

@article{bell2007deep,
  title={Deep-level composition variables as predictors of team performance: a meta-analysis.},
  author={Bell, Suzanne T},
  journal={Journal of applied psychology},
  volume={92},
  number={3},
  pages={595},
  year={2007},
  publisher={American Psychological Association}
}

@article{yu2026towards,
  title={Towards Automated Crowdsourced Testing via Personified-LLM},
  author={Yu, Shengcheng and Ling, Yuchen and Fang, Chunrong and Chen, Zhenyu and Chen, Chunyang},
  journal={arXiv preprint arXiv:2603.24160},
  year={2026}
}

@article{chen2025mimic,
  title={MIMIC: Integrating Diverse Personality Traits for Better Game Testing Using Large Language Model},
  author={Chen, Yifei and Habchi, Sarra and Wei, Lili},
  journal={arXiv preprint arXiv:2510.01635},
  year={2025}
}

@article{ren2025hydra,
  title={Hydra-Reviewer: A holistic multi-agent system for automatic code review comment generation},
  author={Ren, Xiaoxue and Dai, Chaoqun and Huang, Qiao and Wang, Ye and Liu, Chao and Jiang, Bo},
  journal={IEEE Transactions on Software Engineering},
  year={2025},
  publisher={IEEE}
}

@article{li2023large,
  title={Large language models understand and can be enhanced by emotional stimuli},
  author={Li, Cheng and Wang, Jindong and Zhang, Yixuan and Zhu, Kaijie and Hou, Wenxin and Lian, Jianxun and Luo, Fang and Yang, Qiang and Xie, Xing},
  journal={arXiv preprint arXiv:2307.11760},
  year={2023}
}

@article{gnambs2015makes,
  title={What makes a computer wiz? Linking personality traits and programming aptitude},
  author={Gnambs, Timo},
  journal={Journal of Research in Personality},
  volume={58},
  pages={31--34},
  year={2015},
  publisher={Elsevier}
}

@inproceedings{lin2018sentiment,
  title={Sentiment analysis for software engineering: How far can we go?},
  author={Lin, Bin and Zampetti, Fiorella and Bavota, Gabriele and Di Penta, Massimiliano and Lanza, Michele and Oliveto, Rocco},
  booktitle={Proceedings of the 40th international conference on software engineering},
  pages={94--104},
  year={2018}
}

@misc{ONETContentModel, title = {{O*NET} Content Model}, author = {{National Center for O*NET Development}}, year = {2026}, howpublished = {\url{https://www.onetcenter.org/content.html}}, note = {Accessed: 2026-06-14} }

@misc{ONETSoftwareDevelopers, title = {{15-1252.00} -- Software Developers}, author = {{National Center for O*NET Development}}, year = {2026}, howpublished = {{O*NET OnLine}}, url = {https://www.onetonline.org/link/summary/15-1252.00}, note = {Accessed: 2026-06-14} }

@misc{anthropic2026sonnet46, title = {Introducing Claude Sonnet 4.6}, author = {{Anthropic}}, year = {2026}, month = feb, howpublished = {\url{https://www.anthropic.com/news/claude-sonnet-4-6}}, note = {Accessed: 2026-06-14} }

@techreport{anthropic2026sonnet46systemcard, title = {Claude Sonnet 4.6 System Card}, author = {{Anthropic}}, institution = {Anthropic}, year = {2026}, month = feb, url = {https://www-cdn.anthropic.com/78073f739564e986ff3e28522761a7a0b4484f84.pdf}, note = {Accessed: 2026-06-14} }

@inproceedings{
jain2025livecodebench,
title={LiveCodeBench: Holistic and Contamination Free Evaluation of Large Language Models for Code},
author={Naman Jain and King Han and Alex Gu and Wen-Ding Li and Fanjia Yan and Tianjun Zhang and Sida Wang and Armando Solar-Lezama and Koushik Sen and Ion Stoica},
booktitle={The Thirteenth International Conference on Learning Representations},
year={2025},
url={https://openreview.net/forum?id=chfJJYC3iL}
}

@inproceedings{chen-etal-2025-magicore,
    title = "{MA}g{IC}o{R}e: Multi-Agent, Iterative, Coarse-to-Fine Refinement for Reasoning",
    author = "Chen, Justin  and
      Prasad, Archiki  and
      Saha, Swarnadeep  and
      Stengel-Eskin, Elias  and
      Bansal, Mohit",
    editor = "Christodoulopoulos, Christos  and
      Chakraborty, Tanmoy  and
      Rose, Carolyn  and
      Peng, Violet",
    booktitle = "Proceedings of the 2025 Conference on Empirical Methods in Natural Language Processing",
    month = nov,
    year = "2025",
    address = "Suzhou, China",
    publisher = "Association for Computational Linguistics",
    url = "https://aclanthology.org/2025.emnlp-main.1660/",
    doi = "10.18653/v1/2025.emnlp-main.1660",
    pages = "32663--32686",
    ISBN = "979-8-89176-332-6"
}

@inproceedings{qian2025scaling,
  title={Scaling large language model-based multi-agent collaboration},
  author={Qian, Chen and Xie, Zihao and Wang, Yifei and Liu, Wei and Zhu, Kunlun and Xia, Hanchen and Dang, Yufan and Du, Zhuoyun and Chen, Weize and Yang, Cheng and others},
  booktitle={International Conference on Learning Representations},
  volume={2025},
  pages={41488--41505},
  year={2025}
}

@inproceedings{wu2024autogen,
  title={Autogen: Enabling next-gen LLM applications via multi-agent conversations},
  author={Wu, Qingyun and Bansal, Gagan and Zhang, Jieyu and Wu, Yiran and Li, Beibin and Zhu, Erkang and Jiang, Li and Zhang, Xiaoyun and Zhang, Shaokun and Liu, Jiale and others},
  booktitle={First conference on language modeling},
  year={2024}
}

@article{li2023camel,
  title={Camel: Communicative agents for {``Mind''} exploration of large language model society},
  author={Li, Guohao and Hammoud, Hasan and Itani, Hani and Khizbullin, Dmitrii and Ghanem, Bernard},
  journal={Advances in neural information processing systems},
  volume={36},
  pages={51991--52008},
  year={2023}
}

@inproceedings{park2023generative,
  title={Generative agents: Interactive simulacra of human behavior},
  author={Park, Joon Sung and O'Brien, Joseph and Cai, Carrie Jun and Morris, Meredith Ringel and Liang, Percy and Bernstein, Michael S},
  booktitle={Proceedings of the 36th annual acm symposium on user interface software and technology},
  pages={1--22},
  year={2023}
}

@inproceedings{chen2024agentverse,
  title={Agentverse: Facilitating multi-agent collaboration and exploring emergent behaviors},
  author={Chen, Weize and Su, Yusheng and Zuo, Jingwei and Yang, Cheng and Yuan, Chenfei and Chan, Chi-Min and Yu, Heyang and Lu, Yaxi and Hung, Yi-Hsin and Qian, Chen and others},
  booktitle={International Conference on Learning Representations},
  volume={2024},
  pages={20094--20136},
  year={2024}
}

@article{chen2021evaluating,
  title={Evaluating large language models trained on code},
  author={Chen, Mark and Tworek, Jerry and Jun, Heewoo and Yuan, Qiming and Pinto, Henrique Ponde De Oliveira and Kaplan, Jared and Edwards, Harri and Burda, Yuri and Joseph, Nicholas and Brockman, Greg and others},
  journal={arXiv preprint arXiv:2107.03374},
  year={2021}
}

@article{austin2021program,
  title={Program synthesis with large language models},
  author={Austin, Jacob and Odena, Augustus and Nye, Maxwell and Bosma, Maarten and Michalewski, Henryk and Dohan, David and Jiang, Ellen and Cai, Carrie and Terry, Michael and Le, Quoc and others},
  journal={arXiv preprint arXiv:2108.07732},
  year={2021}
}

@article{nijkamp2022codegen,
  title={Codegen: An open large language model for code with multi-turn program synthesis},
  author={Nijkamp, Erik and Pang, Bo and Hayashi, Hiroaki and Tu, Lifu and Wang, Huan and Zhou, Yingbo and Savarese, Silvio and Xiong, Caiming},
  journal={arXiv preprint arXiv:2203.13474},
  year={2022}
}

@article{li2023starcoder,
  title={Starcoder: may the source be with you!},
  author={Li, Raymond and Allal, Loubna Ben and Zi, Yangtian and Muennighoff, Niklas and Kocetkov, Denis and Mou, Chenghao and Marone, Marc and Akiki, Christopher and Li, Jia and Chim, Jenny and others},
  journal={arXiv preprint arXiv:2305.06161},
  year={2023}
}

@article{roziere2023code,
  title={Code llama: Open foundation models for code},
  author={Roziere, Baptiste and Gehring, Jonas and Gloeckle, Fabian and Sootla, Sten and Gat, Itai and Tan, Xiaoqing Ellen and Adi, Yossi and Liu, Jingyu and Sauvestre, Romain and Remez, Tal and others},
  journal={arXiv preprint arXiv:2308.12950},
  year={2023}
}

@article{shinn2023reflexion,
  title={Reflexion: Language agents with verbal reinforcement learning},
  author={Shinn, Noah and Cassano, Federico and Gopinath, Ashwin and Narasimhan, Karthik and Yao, Shunyu},
  journal={Advances in neural information processing systems},
  volume={36},
  pages={8634--8652},
  year={2023}
}

@article{madaan2023self,
  title={Self-refine: Iterative refinement with self-feedback},
  author={Madaan, Aman and Tandon, Niket and Gupta, Prakhar and Hallinan, Skyler and Gao, Luyu and Wiegreffe, Sarah and Alon, Uri and Dziri, Nouha and Prabhumoye, Shrimai and Yang, Yiming and others},
  journal={Advances in neural information processing systems},
  volume={36},
  pages={46534--46594},
  year={2023}
}

@article{huang2023agentcoder,
  title={Agentcoder: Multi-agent-based code generation with iterative testing and optimisation},
  author={Huang, Dong and Zhang, Jie M and Luck, Michael and Bu, Qingwen and Qing, Yuhao and Cui, Heming},
  journal={arXiv preprint arXiv:2312.13010},
  year={2023}
}

@inproceedings{li2022automating,
  title={Automating code review activities by large-scale pre-training},
  author={Li, Zhiyu and Lu, Shuai and Guo, Daya and Duan, Nan and Jannu, Shailesh and Jenks, Grant and Majumder, Deep and Green, Jared and Svyatkovskiy, Alexey and Fu, Shengyu and others},
  booktitle={Proceedings of the 30th ACM joint European software engineering conference and symposium on the foundations of software engineering},
  pages={1035--1047},
  year={2022}
}

@inproceedings{pieterse2006software,
  title={Software engineering team diversity and performance},
  author={Pieterse, Vreda and Kourie, Derrick G and Sonnekus, Inge P},
  booktitle={Proceedings of the 2006 annual research conference of the South African institute of computer scientists and information technologists on IT research in developing countries},
  pages={180--186},
  year={2006}
}

@inproceedings{karn2006follow,
  title={A follow up study of the effect of personality on the performance of software engineering teams},
  author={Karn, John and Cowling, Tony},
  booktitle={Proceedings of the 2006 ACM/IEEE international symposium on Empirical software engineering},
  pages={232--241},
  year={2006}
}

@article{acuna2015team,
  title={Are team personality and climate related to satisfaction and software quality? Aggregating results from a twice replicated experiment},
  author={Acu{\~n}a, Silvia T and G{\'o}mez, Marta N and Hannay, Jo E and Juristo, Natalia and Pfahl, Dietmar},
  journal={Information and Software Technology},
  volume={57},
  pages={141--156},
  year={2015},
  publisher={Elsevier}
}

@inproceedings{islam2024mapcoder,
  title={Mapcoder: Multi-agent code generation for competitive problem solving},
  author={Islam, Md Ashraful and Ali, Mohammed Eunus and Parvez, Md Rizwan},
  booktitle={Proceedings of the 62nd Annual Meeting of the Association for Computational Linguistics (Volume 1: Long Papers)},
  pages={4912--4944},
  year={2024}
}

@inproceedings{bacchelli2013expectations,
  title={Expectations, outcomes, and challenges of modern code review},
  author={Bacchelli, Alberto and Bird, Christian},
  booktitle={2013 35th international conference on software engineering (ICSE)},
  pages={712--721},
  year={2013},
  organization={IEEE}
}

@article{bolker2009generalized,
  title={Generalized linear mixed models: a practical guide for ecology and evolution},
  author={Bolker, Benjamin M and Brooks, Mollie E and Clark, Connie J and Geange, Shane W and Poulsen, John R and Stevens, M Henry H and White, Jada-Simone S},
  journal={Trends in ecology \& evolution},
  volume={24},
  number={3},
  pages={127--135},
  year={2009},
  publisher={Elsevier}
}

@article{bates2015fitting,
  title={Fitting linear mixed-effects models using lme4},
  author={Bates, Douglas and M{\"a}chler, Martin and Bolker, Ben and Walker, Steve},
  journal={Journal of statistical software},
  volume={67},
  pages={1--48},
  year={2015}
}

@inproceedings{yan2025codeif,
  title={Codeif: Benchmarking the instruction-following capabilities of large language models for code generation},
  author={Yan, Kaiwen and Guo, Hongcheng and Shi, Xuanqing and Cao, Shaosheng and Di, Donglin and Li, Zhoujun},
  booktitle={Proceedings of the 63rd Annual Meeting of the Association for Computational Linguistics (Volume 6: Industry Track)},
  pages={1272--1286},
  year={2025}
}

@inproceedings{yu2025humaneval,
  title={HumanEval pro and MBPP pro: Evaluating large language models on self-invoking code generation task},
  author={Yu, Zhaojian and Zhao, Yilun and Cohan, Arman and Zhang, Xiao-Ping},
  booktitle={Findings of the Association for Computational Linguistics: ACL 2025},
  pages={13253--13279},
  year={2025}
}

@inproceedings{yang2025can,
  title={Can LLMs Generate High-Quality Test Cases for Algorithm Problems? TestCase-Eval: A Systematic Evaluation of Fault Coverage and Exposure},
  author={Yang, Zheyuan and Kuang, Zexi and Xia, Xue and Zhao, Yilun},
  booktitle={Proceedings of the 63rd Annual Meeting of the Association for Computational Linguistics (Volume 2: Short Papers)},
  pages={1050--1063},
  year={2025}
}

@article{lin2024awq,
  title={Awq: Activation-aware weight quantization for on-device llm compression and acceleration},
  author={Lin, Ji and Tang, Jiaming and Tang, Haotian and Yang, Shang and Chen, Wei-Ming and Wang, Wei-Chen and Xiao, Guangxuan and Dang, Xingyu and Gan, Chuang and Han, Song},
  journal={Proceedings of machine learning and systems},
  volume={6},
  pages={87--100},
  year={2024}
}

@inproceedings{jiang2024personallm,
  title={PersonaLLM: Investigating the ability of large language models to express personality traits},
  author={Jiang, Hang and Zhang, Xiajie and Cao, Xubo and Breazeal, Cynthia and Roy, Deb and Kabbara, Jad},
  booktitle={Findings of the association for computational linguistics: NAACL 2024},
  pages={3605--3627},
  year={2024}
}

@inproceedings{zhuo2025bigcodebench,
  title={Bigcodebench: Benchmarking code generation with diverse function calls and complex instructions},
  author={Zhuo, Terry Yue and Vu, Minh Chien and Chim, Jenny and Hu, Han and Yu, Wenhao and Widyasari, Ratnadira and Yusuf, Imam Nur Bani and Zhan, Haolan and He, Junda and Paul, Indraneil and others},
  booktitle={International Conference on Learning Representations},
  volume={2025},
  pages={66602--66656},
  year={2025}
}

@article{liu2023your,
  title={Is your code generated by chatgpt really correct? rigorous evaluation of large language models for code generation},
  author={Liu, Jiawei and Xia, Chunqiu Steven and Wang, Yuyao and Zhang, Lingming},
  journal={Advances in neural information processing systems},
  volume={36},
  pages={21558--21572},
  year={2023}
}

@misc{qwen2025qwen25technicalreport,
      title={Qwen2.5 Technical Report}, 
      author={Qwen and An Yang and Baosong Yang and Beichen Zhang and Binyuan Hui and Bo Zheng and Bowen Yu and Chengyuan Li and Dayiheng Liu and Fei Huang and Haoran Wei and Huan Lin and Jian Yang and Jianhong Tu and Jianwei Zhang and Jianxin Yang and Jiaxi Yang and Jingren Zhou and Junyang Lin and Kai Dang and Keming Lu and Keqin Bao and Kexin Yang and Le Yu and Mei Li and Mingfeng Xue and Pei Zhang and Qin Zhu and Rui Men and Runji Lin and Tianhao Li and Tianyi Tang and Tingyu Xia and Xingzhang Ren and Xuancheng Ren and Yang Fan and Yang Su and Yichang Zhang and Yu Wan and Yuqiong Liu and Zeyu Cui and Zhenru Zhang and Zihan Qiu},
      year={2025},
      eprint={2412.15115},
      archivePrefix={arXiv},
      primaryClass={cs.CL},
      url={https://arxiv.org/abs/2412.15115}, 
}

@article{grattafiori2024llama,
  title={The llama 3 herd of models},
  author={Grattafiori, Aaron and Dubey, Abhimanyu and Jauhri, Abhinav and Pandey, Abhinav and Kadian, Abhishek and Al-Dahle, Ahmad and Letman, Aiesha and Mathur, Akhil and Schelten, Alan and Vaughan, Alex and others},
  journal={arXiv preprint arXiv:2407.21783},
  year={2024}
}

@book{cochran1977sampling,
  title={Sampling techniques},
  author={Cochran, William Gemmell},
  year={1977},
  edition={3rd},
  publisher={John Wiley \& Sons}
}

@article{benjamini1995controlling,
  title={Controlling the false discovery rate: a practical and powerful approach to multiple testing},
  author={Benjamini, Yoav and Hochberg, Yosef},
  journal={Journal of the Royal statistical society: series B (Methodological)},
  volume={57},
  number={1},
  pages={289--300},
  year={1995},
  publisher={Wiley Online Library}
}

@inproceedings{papineni2002bleu,
  title={Bleu: a method for automatic evaluation of machine translation},
  author={Papineni, Kishore and Roukos, Salim and Ward, Todd and Zhu, Wei-Jing},
  booktitle={Proceedings of the 40th annual meeting of the Association for Computational Linguistics},
  pages={311--318},
  year={2002}
}

@misc{mistral2025small24b,
  title = {{Mistral-Small-24B-Instruct-2501}},
  author = {{Mistral AI}},
  year = {2025},
  howpublished = {\url{https://huggingface.co/mistralai/Mistral-Small-24B-Instruct-2501}},
  note = {{Hugging Face} model card}
}

@article{arimbur2026many,
  title={How Many Tries Does It Take? Iterative Self-Repair in LLM Code Generation Across Model Scales and Benchmarks},
  author={Arimbur, Johin Johny},
  journal={arXiv preprint arXiv:2604.10508},
  year={2026}
}

@inproceedings{ahmed2017senticr,
  title={SentiCR: A customized sentiment analysis tool for code review interactions},
  author={Ahmed, Toufique and Bosu, Amiangshu and Iqbal, Anindya and Rahimi, Shahram},
  booktitle={2017 32nd IEEE/ACM International Conference on Automated Software Engineering (ASE)},
  pages={106--111},
  year={2017},
  organization={IEEE}
}

@article{mccrae1992introduction,
  title={An introduction to the five-factor model and its applications},
  author={McCrae, Robert R and John, Oliver P},
  journal={Journal of personality},
  volume={60},
  number={2},
  pages={175--215},
  year={1992},
  publisher={Wiley Online Library}
}

@article{scherer2005emotions,
  title={What are emotions? And how can they be measured?},
  author={Scherer, Klaus R},
  journal={Social science information},
  volume={44},
  number={4},
  pages={695--729},
  year={2005},
  publisher={Sage Publications Sage CA: Thousand Oaks, CA}
}

@BOOK{NAP12814,
  author    = {{National Research Council}},
  editor    = "Nancy T. Tippins and Margaret L. Hilton",
  title     = "A Database for a Changing Economy: Review of the Occupational Information Network (O*NET)",
  isbn      = "978-0-309-14769-9",
  doi       = "10.17226/12814",
  url       = "https://nap.nationalacademies.org/catalog/12814/a-database-for-a-changing-economy-review-of-the-occupational",
  year      = 2010,
  publisher = "The National Academies Press",
  address   = "Washington, DC"
}

@article{russell1980circumplex,
  title={A circumplex model of affect.},
  author={Russell, James A},
  journal={Journal of personality and social psychology},
  volume={39},
  number={6},
  pages={1161},
  year={1980},
  publisher={American Psychological Association}
}

@misc{companionArtifact,
  author       = {Ding, Yunyan},
  title        = {{Personality and emotion in Multi-Agent Software Teams}},
  year         = {2026},
  howpublished = {\url{https://github.com/personas-matter-llms/companion-artifact}},
  note         = {Companion repository}
}
\end{document}